\documentclass[journal=jctc,manuscript=article,layout=onecolumn]{achemso}
\usepackage[version=3]{mhchem} 
\usepackage[T1]{fontenc}       
\usepackage{helvet,times}
\usepackage{bm,textcomp, fixltx2e}
\usepackage{amsmath}
\usepackage{graphicx}
\setkeys{acs}{maxauthors= 0}
\usepackage{multimedia}
\usepackage{setspace}
\author{Federico Comitani}
\affiliation[Department of Physics, King's College London]
{Department of Physics, King's College London, Strand, London WC2R 2LS, United Kingdom}

\author{Vittorio Limongelli}
\affiliation[Universit\'{a} della Svizzera Italiana (USI)]
{Universit\'{a} della Svizzera Italiana (USI), Faculty of Informatics, Institute of Computational Science - Center for Computational Medicine in Cardiology, via G. Buffi 13, CH-6900 Lugano, Switzerland}
\alsoaffiliation[University of Naples "Federico II"]
{Department of Pharmacy, University of Naples "Federico II", via D. Montesano 49, I-80131 Naples, Italy}

\author{Carla Molteni}
\email{carla.molteni@kcl.ac.uk}
\affiliation[King's College London]
{Department of Physics, King's College London, Strand, London WC2R 2LS, United Kingdom}

\title[]
  {The free energy landscape of GABA binding to a pentameric ligand-gated ion channel and its disruption by mutations}

\abbreviations{pLGIC: pentameric ligand-gated ion channel, RDL: resistance to dieldrin, GABA: gamma-aminobutyric acid, ECD: extracellular domain,  
MD: molecular dynamics, FM: funnel-metadynamics, CV: collective variable, RMDS: root mean square deviation, FES: free energy surface}

\begin{document}

\begin{abstract}
{Pentameric ligand-gated ion channels (pLGICs) of the Cys-loop superfamily are important neuroreceptors that mediate fast synaptic transmission. They are activated by the binding of a neurotransmitter, but the details of this process are still not fully understood. 
As a prototypical pLGIC, here we choose the insect resistance to dieldrin (RDL) receptor, involved in the resistance to insecticides, and investigate the binding of the neurotransmitter GABA to its extracellular domain at the atomistic level. We achieve this by means of $\mu$-sec funnel-metadynamics simulations, which efficiently enhance the sampling of bound and unbound states by using a funnel-shaped restraining potential to limit the exploration in the solvent. We reveal the sequence of events in the binding process, from the capture of GABA from the solvent to its pinning between the charged residues Arg111 and Glu204 in the binding pocket. We characterize the associated free energy landscapes in the wild-type RDL receptor and in two mutant forms, where the key residues Arg111 and Glu204 are mutated to Ala. Experimentally these mutations produce non-functional channels, which is reflected in the reduced ligand binding affinities, due to the loss of essential interactions. We also analyze the dynamical behaviour of the crucial loop C, whose opening allows the access of GABA to the binding site, while its closure locks the ligand into the protein. The RDL receptor shares structural and functional features with other pLGICs, hence our work outlines a valuable protocol to study the binding of ligands to pLGICs beyond conventional docking and molecular dynamics techniques.

}
\end{abstract}

\section*{Introduction}

Pentameric ligand-gated ion channels (pLGICs) are important neuroreceptors involved in fast synaptic communication and many neurological disorders. \cite{Lemoine2012}
They are targets for drugs and, in invertebrates, for insecticides.
They are membrane proteins, composed of five subunits arranged around an ion permeable channel, with an extracellular domain (ECD), a transmembrane domain (TMD) spanning the cell membrane and, often, 
an intracellular domain (ICD). 
The channel opens in response to the binding of a neurotransmitter at the interface between subunits in the ECD. 
However a detailed picture at the atomistic level of the activation mechanisms is still missing, due to the complexity of the systems and the limited experimental information. 

Computational studies complement the picture coming from experiments. 
Spectroscopic techniques and/or homology modelling provide the atomistic structure of these receptors' ECD,
which contains the relevant region for ligand binding. This information has
been used in docking and molecular dynamics (MD) calculations to study the
binding modes of ligands to pLGICs. Unfortunately, the approximate description
of ligand/protein interactions by docking algorithms and the limited timescale of MD simulations allow
only a partial understanding of the binding process. 
A comprehensive elucidation of the
activation mechanisms in pLGICs needs detailed information of the ligand
binding steps 
and an accurate description of the free energy landscape, which remain
elusive to standard simulation techniques and require the use of more sophisticated calculations. 
\cite{Michel2010} 

Here we use the well-tempered metadynamics method
\cite{Barducci2008}, which facilitates the sampling through the introduction of a history dependent bias potential. Such a repulsive potential is built as a sum of Gaussians deposited along the trajectory of carefully selected slowly varying degrees of freedom, named collective variables (CVs). Its role is to discourage the system to return
to already visited regions in the CVs space.
This allows to greatly improve the exploration 
of the phase space and, once the exploration is complete, to efficiently reconstruct the free energy landscape as a function of the CVs using the added bias.
Metadynamics has been used to successfully investigate numerous problems in biology, chemistry and material sciences. \cite{Laio2008,
Ensing2006, Melis2009, Leone2009, 
DiLeva2014, 
Cavalli2015}
However, its application to estimate binding free energies is far from trivial. 
For this goal, the observation of a statistically significant number of forth-back events between the ligand bound and unbound states is necessary. Within the metadynamics scheme, while a ligand can be easily driven out of its protein binding site, the huge number of unbound conformations
in the solvent dramatically decreases the probability to observe a binding event in a reasonable computational time.
A simple and elegant solution to this problem is the introduction of a funnel-shaped restraining potential to limit the exploration of the
solvated states outside the binding pocket.\cite{Limongelli2013} 
The funnel is placed onto the system so to include the binding site in the conical section and reduce
the solvated region to a narrow cylinder.
In this way, the sampling in the binding site is not affected by the presence of the funnel restraining potential and in this region
the simulation proceeds as standard metadynamics.
On the other hand, the presence of the cylinder restraint in the unbound region reduces the phase space exploration. As a result, the number of forth and back events between the bound
and the unbound state is greatly enhanced \cite{Limongelli2013, Bruno2015, Hsiao2014, Troussicot2015}, leading to an accurate description of the binding
free energy landscape in an affordable computational time.

In the present work, we apply for the first time this funnel-metadynamics (FM) scheme to investigate the free energy landscape of a neurotransmitter binding to a pLGIC and assess how such a landscape is affected by specific mutations, which experimentally produce non-functional channels.
Rationalizing the different binding mechanisms and affinities of a ligand 
towards the wild-type and mutant forms of its molecular target is, in general,  
a crucial issue in the chemical and pharmacological industries.

As a prototypical example, we choose the insect resistance to dieldrin (RDL) $\gamma$-aminobutyric acid  (GABA)-gated pLGIC, from the Cys-loop receptors superfamily. 
This choice is driven by the wealth of the available experimental information \cite{McGonigle2010, Ashby2012}
and by our expertise (modelling and MD simulations) on the system. \cite{Ashby2012, Comitani2014}
The RDL receptor is involved in the resistance to insecticides (i.e. dieldrin) and is thus a potential target for the rational design of novel insecticides.  
\cite{Rahman2014, Liu2015, Liu2015a}

\begin{figure*}
\includegraphics[width=17.8cm]{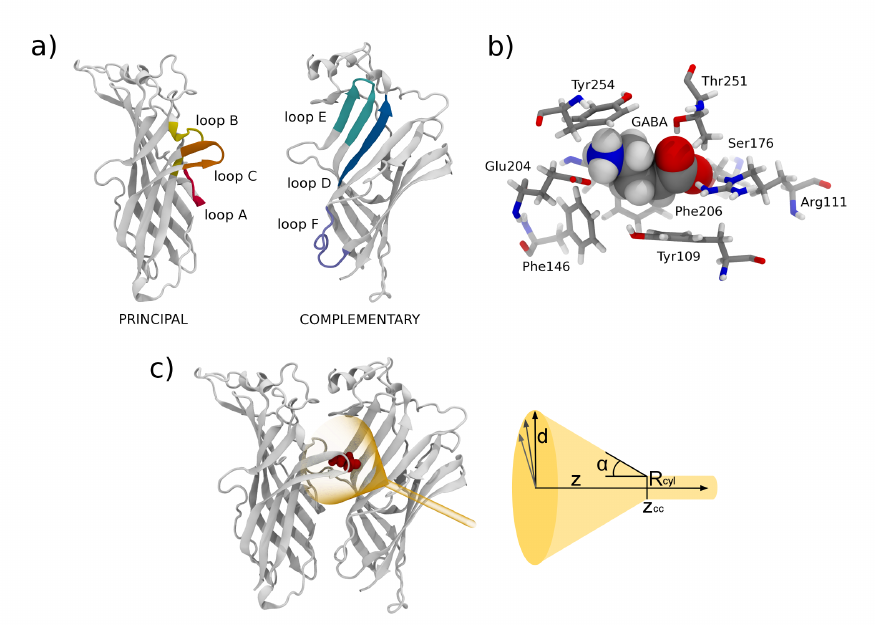}
\caption
{
a) Principal and complementary subunits of the RDL receptor ECD,
with highlighted secondary structures.
b) GABA in the binding site with the seven residues identified by experiments as important for binding and Thr251 in loop C.
c) Left, principal and complementary subunits with GABA bound in red and the funnel restraining potential. Right, geometric parameters of the funnel restraining potential.} 
\label{fig:fig1}
\end{figure*}

Most experimental information on GABA binding to the RDL receptor is indirectly inferred from electrophysiology mutagenesis experiments,
which identified seven residues as important for ligand binding: Phe146, Glu204, Phe206 and Tyr254 in the principal subunit of the RDL receptor ECD, and Tyr109, Arg111 and Ser176 in the complementary subunit (see Figure \ref{fig:fig1}). \cite{Ashby2012} 
Our previous atomistic MD simulations showed that the zwitterionic form of GABA interacts with the positively charged Arg111 through its negatively charged carboxylate and with the negatively charged Glu204 through its positively charged amine, which also forms cation-$\pi$ interactions with the aromatic residues  
Tyr109, Tyr254 and Phe206.\cite{Comitani2014, Comitani2015}
These results are consistent with the experimental data showing that mutations of the Arg111 and Glu204 to Ala produced non-functional receptors. \cite{Ashby2012} However, due to the limited timescale of those simulations, the binding mechanism of GABA to the RDL receptor and a reliable estimate of the ligand binding free energy landscape could not be provided. 

Here, we complete the picture stemming from experiments and previous simulations of the binding process of GABA to the RDL receptor and its non-functional Arg11Ala and Glu204Ala mutants. With $\mu$-sec FM simulations, we reproduce a large number of binding and unbinding events that lead to a quantitatively well-characterized free energy landscape and accurate estimate of the ligand binding free energy.  
Furthermore, we provide structural insights into the ligand binding mechanism with atomistic resolution, identifying crucial protein motions 
like the dynamics of loop C (highlighted in 
Figure \ref{fig:fig1}),
which partially protects the binding site from the solvent and is expected to play an important role in the activation processes of Cys-loop receptors. \cite{Cheng2006, Lee2009, Yakel2010}

Since the RDL receptor is structurally similar to other pLGICs, our work represents a point of reference for further investigations of the binding of neurotransmitters and ligands to pLGICs beyond conventional docking and MD-based techniques.

\section*{Methods}

The extracellular domain of the homopentameric RDL receptor
is built with the software MODELLER 9.8 \cite{Eswar2006} 
by homology with the x-ray structure of the GluCl receptor, \cite{Hibbs2011} 
with which it shares 38$\%$ of sequence identity.\cite{Ashby2012, Comitani2014}
The zwitterionic form of GABA is docked in it with the software GOLD \cite{Verdonk2003}
and the guide of mutagenesis experiments (model RDL-GluCl-2b in Ref. \cite{Comitani2014}); the initial docking pose is not crucial for the metadynamics simulations. 
We have previously extensively tested this model, also in comparison to others; we select it for this work for its stability and binding features consistent with experiments.
We concentrate on the binding and unbinding of a single ligand; MD simulations with one or five ligands did not provide significant differences in the binding features. \cite{Comitani2014}
We simulate the ECDs of the wild-type RDL receptor and of the Arg111Ala and Glu204Ala mutants. In our models, as in previous work, we do not include 
the TMD and lipid membrane since they do not play an active role during the binding process: the binding pocket is far away from them and the slow conformational changes initiated by neurotransmitter binding and then transmitted to the TMD are not sampled in these simulations.
Models with only the ECD are commonly used to study the binding process in ligand-gated ion channels, \cite{Melis2008,Ashby2012,Comitani2014} also due to the limited structural information for the complete receptors. The reduced models allow us to afford longer time scales while retaining all the important features for the GABA binding process. 
The C$_\alpha$s of the five terminal residues of each subunit are restrained 
to mimic the effect of the TMD at the interface with the ECD. 

Following a similar protocol to Ref. \cite{Comitani2014} , we 
prepare the system with a 12 \mbox{\normalfont\AA}\  buffer of TIP3P waters and 0.15 M 
(NaCl) salinity to reproduce 
physiological conditions in a periodically repeated truncated octahedral supercell; counter ions are added to neutralise the total charge.
The solvated model contains about 83,000 atoms.
The AMBER FF12SB \cite{Case2012} force field is used; this is an improved force field with respect to the FF2003 we previously employed.
\cite{Comitani2014, Melis2008} Moreover, with respect to the widely used FF99SB-ildn which we also tested, it provides a better description on long time scales of loop C, which is critical for the binding process. 
Restrained ESP charges are calculated for GABA at the Hartree Fock level of theory, with a 6-31G* basis set, on the molecular geometry optimised 
with density functional theory and the B3LYP exchange and correlation functional using Gaussian 09 \cite{Frisch2009} and 
AmberTools.\cite{Case2012}
All metadynamics simulations are run with NAMD 2.9 \cite{Phillips2005} patched with the PLUMED 1.3 metadynamics plug-in. \cite{Bonomi2009a} 
To enforce ambient conditions (T=300 K and P=1 bar), the Langevin thermostat, with a collision frequency of 1 ps$^{-1}$,
and the Berendsen barostat, with relaxation time of 0.2 ps, 
are applied. A cutoff of 10 \mbox{\normalfont\AA}\ is used for non-bonded interactions and long distance electrostatic interactions 
are treated with Particle-Mesh Ewald. 
The fast stretching motions of bonds containing hydrogen are prevented by means of the  
SHAKE algorithm,
allowing for a 2 fs timestep. 
After the minimisation and equilibration procedures, 
100 ns of MD are carried out, during which the stability of the structure is monitored through the root mean square deviation 
(RMSD) of the backbone atoms of the secondary structures with respect to the minimised structure, 
before starting the metadynamics simulations. 

A number of exploratory metadynamics simulations are first performed without the restraining funnel potential to get an idea 
of the preferred path for unbinding with statistical significance. 
Gaussians with a height of 0.4 kJ/mol and a width of 
0.15 \mbox{\normalfont\AA}\ 
are
deposited every 2 ps along the trajectory of the selected CV: the distance between the center of mass of 
GABA and the center of mass of the C$_\alpha$s of the seven residues identified by experiments as important for binding. 
This procedure allows us to study the 
behaviour of the secondary structures surrounding the binding site upon unbinding. 
We observe that loop C is able to open and close quickly enough following the passage of the ligand, when described with the FF12SB force field;
thus it does not represent an obstacle to the binding of the ligand and does not require a specific CV to describe its motion. 
The information gathered from these preliminary runs allows us to position the funnel-shaped restraining potential for GABA center of mass in a meaningful way
as shown in Figure \ref{fig:fig1} c): the position of the conical region surrounding the binding site is chosen 
so not to interfere with the binding and unbinding processes, while its dimensions are chosen as 
reduced as possible to speed up convergence, but large enough to avoid artifacts in the sampling. 
The cone angle $\alpha$ is set to 25$^\circ$. The zero of the coordinate system is set around the minimum free energy for the wild-type receptor, with the conical region between  ${\rm z=-5}$\ \mbox{\normalfont\AA}\ and ${\rm z_{cc} = 20 }$\ \mbox{\normalfont\AA}, the point where  the potential switched to a cylindrical shape with radius ${\rm R_{cyl} =1 }$\ \mbox{\normalfont\AA}\  
and length of 10\ \mbox{\normalfont\AA}. Soft harmonic restraining walls are applied at both ends of the funnel axis to limit the exploration 
of the ligand inside the region enclosed by the funnel potential.

The zwitterionic form of the ligand and the presence of residues of opposite charge in the binding site (i.e. 
Glu204 and Arg111 in the principal and complementary subunits respectively)
support the picture that 
plausible binding poses should all have a similar orientation with respect to the binding pocket, with GABA carboxylate group interacting with Arg111 and the amine group interacting with Glu204, making it unnecessary to use a CV describing the orientation of GABA. 
This is consistent with 
the unbinding tests, where no major 
rotations or change in orientation were observed when the ligand was inside the binding pocket. 
Hence, 
as CVs for the metadynamics, we choose the position along the funnel axis (${\rm z}$) and the distance from it (${\rm d}$), as shown in Figure \ref{fig:fig1} c): these are an appropriate selection as they allow a complete exploration of the space inside the funnel potential
without producing hysteresis in the binding and unbinding paths or other artifacts.
Although these CVs easily discriminate between bound and unbound states, a reweighting algorithm \cite{Bonomi2009} is applied to the trajectories in order to remap the free energy as a function of alternative CVs and gain more detailed insights, such as 
revealing bound minima which are overlying in the maps corresponding to the initial choice. The results obtained by such algorithm need to be properly analysed to make sure that the unbiased CVs chosen for the reweighting have been adequately sampled during the simulation. When this is not the case, partial or incorrect reconstructions may occur.

Gaussians are deposited every 2 ps, under the well-tempered regime. \cite{Barducci2008} The initial height is
2.0 kJ/mol, the target temperature 300 K and the bias factor 15; the simulations are carried out in the NVT ensemble starting with the average volume obtained in the equilibrated 
MD. 
The calculations are considered converged when a statistically significant number of re-crossing events between the bound and unbound states, defined as those conformations
contained in the cylindrical section of the funnel potential, have been observed and the difference in free energy
between these (bound and unbound) states oscillates around a constant value. 
In the wild-type receptor the metadynamics simulation are carried out for 1.25 $\mu$s.  
The same protocol is then applied to the two Arg111Ala and Glu204Ala mutants, with the goal to compare the resulting 
free energy landscapes.
The metadynamics simulations are carried out for 
0.75 $\mu$s for the Arg111Ala RDL receptor and for 0.95 $\mu$s for the Glu204Ala RDL receptor, where the binding is weaker and the ligand more likely to exit.
The free energy landscapes and binding free energies are averaged over the last 0.25 $\mu$s in all models, to reduce the metadynamics error and increase the accuracy of the reconstructed profiles.\cite{Micheletti2004, Laio2008} Binding affinities are evaluated following the approach described in Ref. \cite{Limongelli2013}, which accounts for the effects of the cylindrical restraint on the solvated states. 

\section*{Results and Discussion}

During the metadynamics simulations we monitor the conformational stability of the RDL receptor models by calculating the RMSD of the secondary structure backbone atoms with respect to the corresponding initial minimized structures. The average RMSD is 2.3$\pm$0.2\ \mbox{\normalfont\AA}\ for the wild-type RDL receptor,
2.6$\pm$0.2 \mbox{\normalfont\AA}\
for the Arg111Ala mutant and 2.6$\pm$0.1 \mbox{\normalfont\AA}\ for the Glu204Ala mutant. 
Hence the overall structure of the receptor is preserved throughout the $\mu$-sec long simulations, during which the
loops and the binding site residues show the necessary flexibility to allow the binding and unbinding of the neurotransmitter.

\begin{figure*}[h!]
\includegraphics[width=17.8cm]{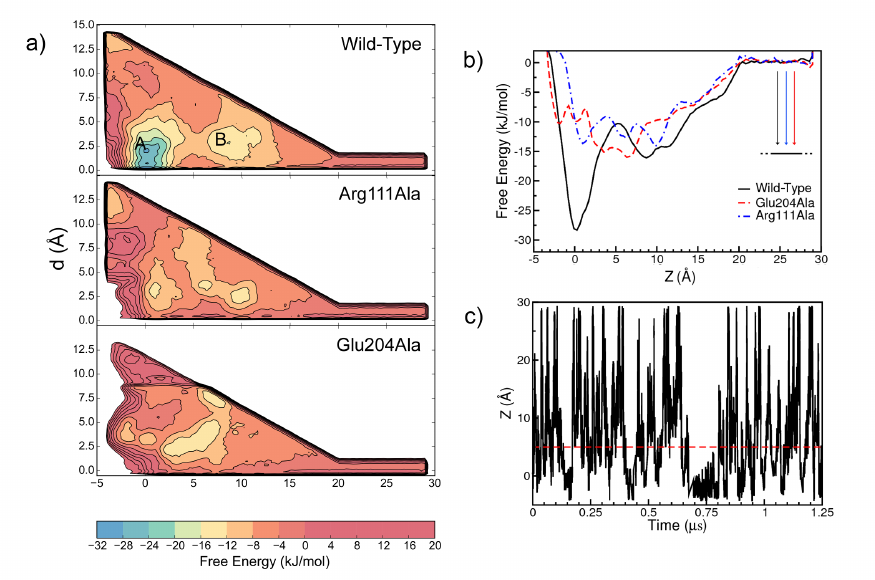}
\caption
{
a) Free energy maps as a function of the biased CVs for GABA binding to the wild-type and the mutated Arg111Ala and Glu204Ala RDL receptors. The free energy of the unbound state is 
set as reference to zero. All values of free energy above zero are shown in dark red. b) Projected free energy profiles as a function of the position ${\rm z}$ along the funnel. The correction due to the cylindrical part of the funnel potential which lowers the free energy of the unbound states is shown and indicated with vertical arrows. 
c) Evolution of the position ${\rm z}$
of GABA center of mass along the funnel axis during the wild-type receptor metadynamics. 
The red dashed line indicates the 
separation between bound and unbound states.}
\label{fig:fig2}
\end{figure*}

In the free energy surfaces (FESs) as a function of the biased CVs in Figure \ref{fig:fig2} a) 
it is clear that the selected mutations induce drastic changes in the free energy landscape. This is even more evident in the one-dimensional free energy profiles projected onto the funnel axis in Figure \ref{fig:fig2} b): an energetically favourable bound state for GABA is present only in the wild-type receptor, which we first analyze in the following.
 
In the wild-type receptor the lowest free energy pose corresponds to the minimum free energy basin A in the FES of Figure \ref{fig:fig2} at
 ${\rm z} \simeq 0$ \mbox{\normalfont\AA}\ and $0 < {\rm d} < 2$ \mbox{\normalfont\AA}\ and to the global minimum at ${\rm z}\simeq 0$ \mbox{\normalfont\AA}\ in the one-dimensional free energy profile. 
Panel c) in Figure \ref{fig:fig2} shows the evolution of the projection of the center of mass of GABA onto the funnel axis during the FM simulation. 
Several binding (${\rm z}<5$ \mbox{\normalfont\AA}) and unbinding events are reproduced, ensuring an exhaustive exploration of the phase space and a quantitatively well-characterized FES.   
The absolute ligand binding affinity is estimated by computing the free energy difference between the bound state in basin A and the isoenergetic states in the unbound region (with ${\rm z} > 20$ \mbox{\normalfont\AA}). The calculated value 
is $-29.7 \pm 2.9$ kJ/mol. However, the free energy of the unbound states has to be corrected to remove the effects of the restraining potential\cite{Limongelli2013} (which does not affect the free energy of the bound state), reducing the true binding affinity to
$-14.4 \pm 2.9$ kJ/mol. The entity of the correction to the unbound states due to the cylindrical part of the funnel potential is shown in Figure \ref{fig:fig2} b).

\begin{figure*}[h!]
\includegraphics[width=8.7cm]{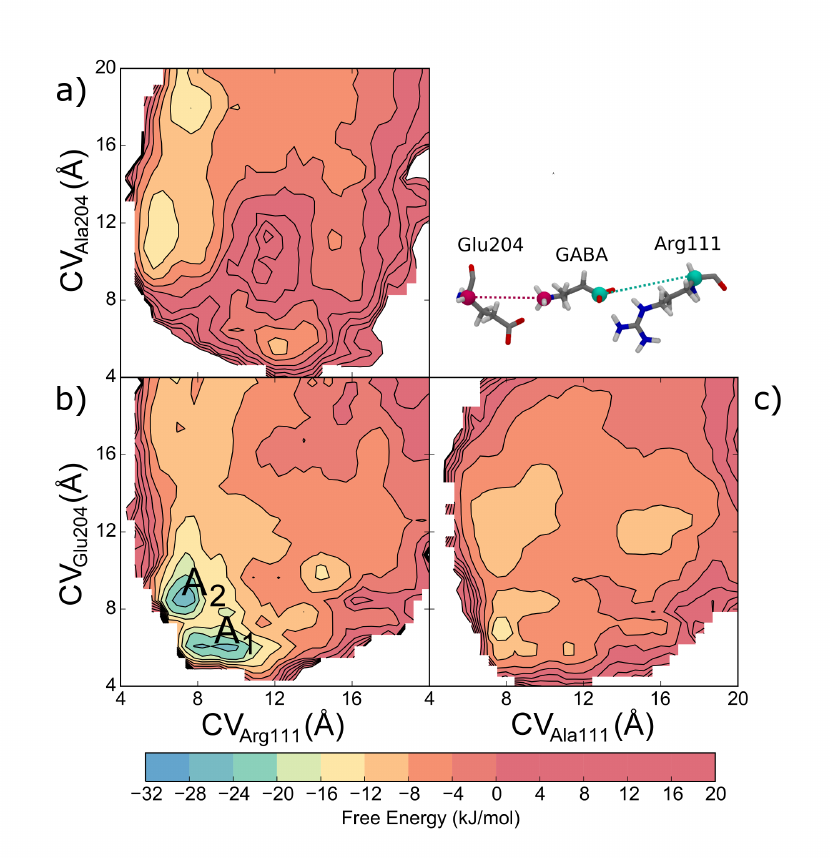}
\caption
{
Free energy maps of a) the Glu204Ala-mutated RDL receptor, b) the wild-type RDL receptor and c) the Arg111Ala-mutated RDL receptor reweighted as a function of the distances between GABA charged groups and the C$_\alpha$s of the 111 and 204 residues. These distances are sketched in the top right corner in cyan and magenta.
}
\label{fig:fig3}
\end{figure*}

\begin{figure*}[h!]
\includegraphics[width=17.8cm]{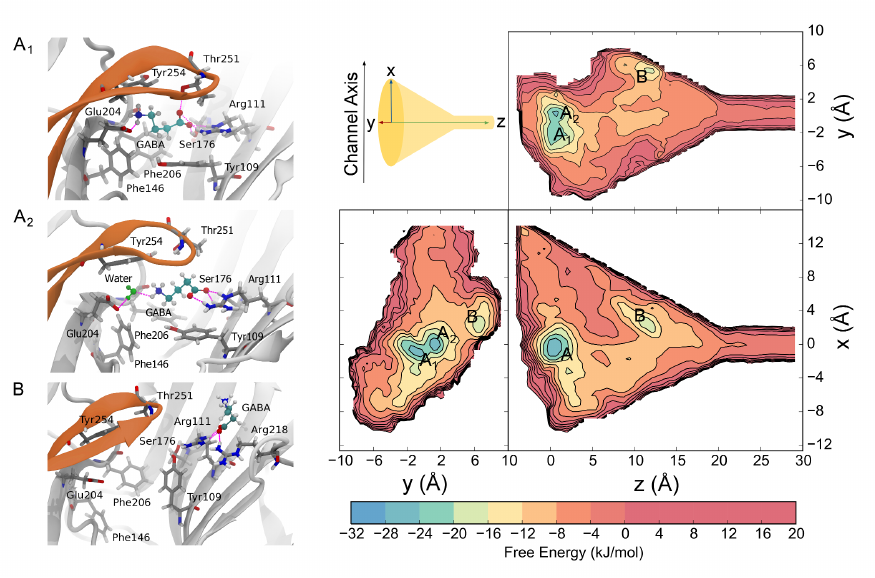}
\caption
{
Right, binding free energy maps in the wild-type RDL receptor projected onto the funnel sections. 
Left, typical conformations from the identified free energy basins; GABA is represented in CPK style, the C-loop is in orange and hydrogen bonds are indicated in magenta. 
In A$_2$ the water molecule that mediates the hydrogen bond interaction between GABA and Glu204 is represented in CPK style and in green.
}
\label{fig:fig4}
\end{figure*}

To provide further structural details on the ligand binding, we remap the FESs as a function of different CVs with respect to those originally biased in the FM simulations, 
using a reweighting algorithm. \cite{Bonomi2009}.
In Figure \ref{fig:fig3} b) the wild-type RDL receptor FES is shown as a function of the distance between the carbon of GABA carboxylate group and the C\textsubscript{$\alpha$} of Arg111 (indicated as ${\rm CV_{Arg111}}$) and of the distance between the nitrogen of GABA amino moiety and the C\textsubscript{$\alpha$} of Glu204 (${\rm CV_{Glu204}}$). 
The reweighted FES reveals two distinct energy minima in the bound region corresponding to two different binding modes, 
here labeled as A\textsubscript{1} and A\textsubscript{2} respectively, which are overlapped in the same basin A
in the FES representation of Figure \ref{fig:fig2}.
We monitor the evolution of ${\rm CV_{Arg111}}$ and ${\rm CV_{Glu204}}$ during the simulation, observing that the two basins of minimal free energy are visited several times. This behavior indicates that the two CVs are adequately sampled even if not biased, making the reweighted FES reliable.
The two minima are also observed in the reweighted FES  
as a function of the position of GABA with respect to the x,y,z Cartesian axes in Figure \ref{fig:fig4}, where z and x are the axes of the funnel and of the channel respectively. 
Here the two minima A\textsubscript{1} and A\textsubscript{2} have coordinates 
${\rm x}\simeq -1$ \mbox{\normalfont\AA}, ${\rm y}\simeq -2$ \mbox{\normalfont\AA}, ${\rm z} \simeq 4$ \mbox{\normalfont\AA}\ and ${\rm x}\simeq 0$ \mbox{\normalfont\AA}, ${\rm y} \simeq 0$ \mbox{\normalfont\AA}, ${\rm z}\simeq 4$ \mbox{\normalfont\AA}\ respectively. 
As seen in Figure \ref{fig:fig2}, Figure \ref{fig:fig3} and Figure \ref{fig:fig4},
the details of the basins, transition states and barriers may depend on the choice of the CVs used to represent the FES.  

We perform a conformational cluster analysis of the poses representing the two minimal free energy basins A\textsubscript{1} and A\textsubscript{2}.
In A\textsubscript{1}, the most populated family corresponds to the  
binding mode where GABA forms strong direct hydrogen bonds with both Arg111 and Glu204
as in Figure \ref{fig:fig4}. Specifically,
the negatively charged carboxylate moiety of GABA forms salt bridges with Arg111 in loop D and interacts through multiple hydrogen bonds with Ser176 in loop E and 
Thr251 in loop C. 
The latter loop is conformationally rather flexible during the simulation, allowing the access and the exit of the ligand to and from the binding pocket 
through its closure and opening motion.  
GABA amine is bound, through hydrogen bonds, to the residues of loop B Glu204 and Ser205, and can also form hydrogen bonds with the hydroxyl groups of Tyr109 in loop D and Tyr254 in loop C. The amine is buried in an aromatic cage formed by these two tyrosines and Phe206 in loop B  with which it forms cation-$\pi$ interactions. 
Consistently with this picture, in Figure \ref{fig:fig3} (b), basin A\textsubscript{1} is located at  ${\rm CV_{Arg111}} \simeq 8$ \mbox{\normalfont\AA}\ and ${\rm CV_{Glu204}} \simeq 6$ \mbox{\normalfont\AA}, with GABA tightly bound to both the primary and the complementary subunit. 

The A\textsubscript{2} basin is observed at ${\rm CV_{Arg111}} \simeq 7$ \mbox{\normalfont\AA}\ and ${\rm CV_{Glu204}} \simeq 8$ \mbox{\normalfont\AA}\ in Figure \ref{fig:fig3} (b).
The cluster analysis of the poses corresponding to this basin shows, in the most populated family, GABA bound to the residues of the complementary unit, i.e. Arg111, Ser176, and to Thr251 in loop C. 
Similarly to pose A\textsubscript{1}, GABA amine group is held in place by cation-$\pi$ interactions with residues of the aromatic cage formed by Tyr109, Tyr254 and Phe360 and by hydrogen bonds with the hydroxyl group of the two tyrosines and Ser205. 
However, at variance from A\textsubscript{1}, in this basin the interaction with Glu204 is typically mediated by a water molecule present in the binding site and coloured in green in Figure \ref{fig:fig4}. 
As a consequence, GABA assumes a slightly different position within the binding pocket with respect to 
the A\textsubscript{1} basin, which allows the concurrent presence of a water molecule without altering the main interactions formed by the ligand with both subunits.
Our results  
highlight the key role played by waters during the binding process, confirming 
the necessity of explicitly including water molecules in the simulations. 
Water mediated interactions have been suggested for other Cys-loop receptors and analogous systems.
\cite{Amiri2007, Blum2010} 
In this representation the difference in free energy between the two minima is $\simeq$ 1\ kJ/mol, with the A\textsubscript{1} minimum slightly favoured;
the two states are easily interchangeable at physiological conditions.

A shallow energy minimum, B, is found in the FES in Figure \ref{fig:fig2} at 
7.5 \mbox{\normalfont\AA} < ${\rm z}$ <12 \mbox{\normalfont\AA}\ 
and 1 \mbox{\normalfont\AA}\ $< {\rm d} <$ 5 \mbox{\normalfont\AA}, farther from the binding site and closer to the cylinder region of the funnel restraint.
This minimum represents a pre-binding pose, as also
suggested by the one-dimensional free energy profile in  Figure \ref{fig:fig2}. 
We stress that this free energy minimum is fully within
the cone section of the funnel, where no external potential is applied to the system and therefore it is not
an artifact.
The conformational cluster analysis of the poses populating this basin
shows a number of different states where GABA is almost out of the binding site,
but still able to interact with the receptor through its carboxylate moiety. 
Here the GABA carboxylate group forms hydrogen bonds with Arg111, whose side chain is tilted outward with respect to the channel axis,
and Arg218, while its amine group points towards the solvent. 
As it can be seen in the left bottom panel of Figure \ref{fig:fig4}, in this state loop C is slightly open and no interaction is observed between this loop and the complementary subunit.  
Basin B is 12.4 kJ/mol higher in free energy than basin A.
From this pose the ligand can reach the lowest energy state A, which corresponds to the final binding mode,
crossing a small energy barrier as evident in Figure \ref{fig:fig2}.
During the passage from the pre-binding mode B to the binding mode A, the ligand interacts with the residues of loop C:
the subsequent closure of the loop locks the ligand in the binding pocket. 

\begin{figure*}
\centering
\includegraphics[width=8.7cm]{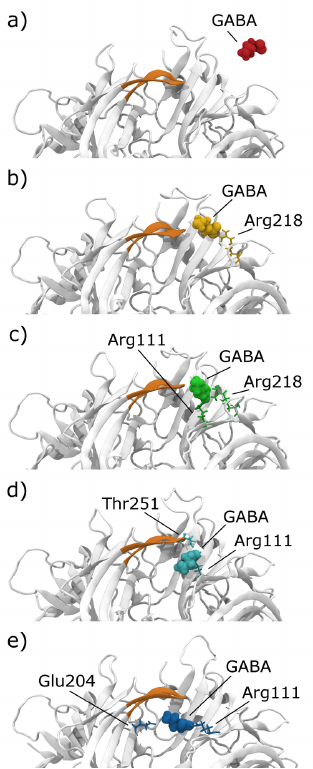}
\caption 
{
Snapshots of a favourable binding path observed in the metadynamics simulations for the wild-type RDL receptor, here seen from the membrane orthogonally to the channel axis. 
GABA and the residues interacting with it through
hydrogen bonds are shown in colours that go from red (at the relative metadynamics time of 0 ns, a), through yellow (1.4 ns, b), green (1.8 ns, c) and cyan (2.7 ns, d), to blue (9.3 ns, e). Loop C is 
in orange.
}
\label{fig:fig5}
\end{figure*}

The sequence of the events characterizing the binding path of GABA is shown in Figure \ref{fig:fig5}. For the sake of clarity the RDL receptor ECD is here displayed perpendicular to the membrane surface.
The corresponding movie is available as Supplementary Information.
In the wild-type RDL receptor, the first protein residue that interacts with GABA, when it is still almost fully solvated, is Arg218 in the complementary subunit. 
Its conformational flexibility and the positive charge of its side chain are crucial to 
capture
the negatively charged carboxylate group of GABA from the solvent.
The GABA-Arg218 interaction is maintained until the formation of hydrogen bonds between GABA and the side chain of
Arg111, which also belongs to the complementary subunit. 
Thus, the GABA carboxylate is exchanged from Arg218 to Arg111,
which acts as the first attractor of the ligand into the binding site thanks to the flexibility and charge of its side chain. 
These interactions with the complementary subunit are instrumental in optimizing the orientation of GABA with respect to the receptor binding pocket.
In this phase, loop C has a fairly open conformation that
allows the ligand to approach the binding pocket. 
The entry of the ligand into the binding site is also facilitated by a hydrogen bond interaction formed by the GABA carboxylate with the hydroxyl group of Thr251 in loop C.
The carboxylate group of GABA is the main player during this preliminary phase of the binding process, while the amine group does not actively take part in the action, remaining in an almost fully solvated state.
To reach the final binding state, GABA passes under loop C, preserving, however, its interactions with Arg111 and Thr251.
Upon the ligand passage, the loops closes,
dragged by 
the ligand contact with Thr251, while  
the GABA amine moiety is positioned so to optimally interact with Glu204, 
through either direct or water-mediated hydrogen bonds, and with the surrounding aromatic residues, through cation-$\pi$ interactions.

\begin{figure*}[h!]
\centering
\includegraphics[width=8.7cm]{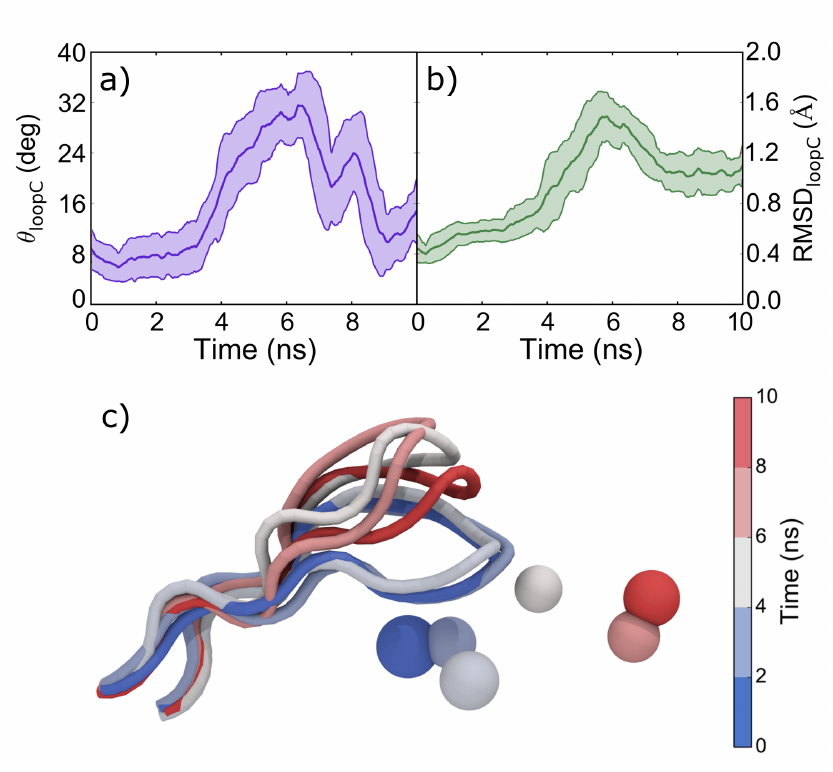}
\caption
{
Evolution of a) the angle describing the orientation of loop C  and b) of the RMSD of  loop C during GABA first unbinding event in the wild-type RDL receptor. 
In panel c) the structure of loop C and the center of mass of GABA, represented as a sphere, are shown every 2 ns with blue indicating the bound state and red the unbound state.
}
\label{fig:fig6}
\end{figure*}

The conformational changes of loop C play a key role also in the unbinding of the ligand from the binding site.
In fact, the opening of loop C favours the release of GABA followed by a partial reclosure of the loop after the ligand unbinding.
This mechanism is clearly demonstrated
by the time evolution of the angle ($\theta_{\rm loopC}$) defined by the axis of loop C and of the RMSD of the 246-255 residues C$_\alpha$s (around the apex of loop C)  at the start and during the simulation.
Their running averages over subsets of one thousand points and the corresponding standard deviations are shown 
in Figure \ref{fig:fig6} a) and b) respectively
for the first unbinding event observed in the FM simulation; panel c) shows the motion of the loop and the relative position of GABA during the unbinding.
The full closure of loop C occurs only when GABA is in the binding pocket and interacts with Thr251, as previously described. 

We now analyze the effects of two key mutations Arg111Ala and Glu204Ala on the binding of GABA to the RDL receptor.

For the Arg111Ala-mutated RDL receptor,   
a few shallow minima appear in the region $0< {\rm z} < 12$ \mbox{\normalfont\AA}\ and $0 < {\rm d} < 5$ \mbox{\normalfont\AA}\ of the FES and in the corresponding one-dimensional free energy profile in Figure \ref{fig:fig2}.
The lower depth of these minima with respect to those of the wild-type receptor is also evident
in the FES representation of Figure \ref{fig:fig3} (a and c).
No deep free energy minimum is present, thus suggesting a poor affinity of GABA for this mutant form.
When considering the entropic cost of using the funnel potential, these data are consistent with weak or no binding in agreement with mutagenesis experiments which recorded non-responsive channels. 
\cite{Lummis2011,Ashby2012}
From the trajectory analysis, we observe that, when the ligand approaches to the Arg111Ala-mutated RDL receptor, the first point of interaction is still Arg218, but the absence of Arg111 prevents GABA from optimally entering the binding pocket and forming strong interactions with its residues. Nevertheless, the ligand manages to
form hydrogen bonds through its carboxylate group with the hydroxyl group of Thr88, and through its amine group with Thr250 and Thr251 in loop C, but no cation-$\pi$ interactions. These interactions give rise to the shallow minimum at $5 < {\rm z} < 12$ \mbox{\normalfont\AA}\ in the FES 
of Figure \ref{fig:fig2}. 
Another shallow energy minimum is found around  ${\rm z} \simeq 1.5$ \mbox{\normalfont\AA}\ and ${\rm d} \simeq 3.0$ \mbox{\normalfont\AA}. Here GABA is closer to the wild-type binding site,
but the reduced and weaker interactions with the protein makes this state less stable than the corresponding pose in the wild-type receptor.
In this basin, Glu204 represents the anchor point of GABA in the principal subunit, where the amine group forms cation-$\pi$ interactions with Phe206 and Tyr254. The latter is also able to engage in sporadic hydrogen bonding with the ligand. In the complementary subunit, Ser176  and Tyr109 interact with the GABA carboxylate, but
these residues are farther apart from Glu204 than position 111, thus forming weaker interactions.
Loop C is slightly more flexible 
if compared with the wild-type receptor and is
unable to close completely and to lock the ligand into the binding pocket. 

Similarly to what was observed for the Arg111Ala mutant, the mutation of Glu204 in loop B with alanine causes 
significant changes in the binding mechanism of GABA with altered interactions with the protein.
The FES in Figure \ref{fig:fig2} shows no energy minimum around ${\rm z}\simeq 0$ \mbox{\normalfont\AA}\ and ${\rm d} \simeq 0.5$ \mbox{\normalfont\AA}, thus indicating that in this form GABA is not able to bind in the
binding pocket identified in the wild-type.
A wide shallow free energy basin
is observed 
at $3.0<{\rm z}<7.5$ \mbox{\normalfont\AA}\ in Figure \ref{fig:fig2} and 
its depth is consistent with very weak or no binding. 
In this basin the GABA carboxylate is bound to the residues of the complementary subunit Arg111 and Arg218, while the amine group does not form any persistent interaction.
The ligand forms hydrogen bonds in the principal subunit with Thr251 in loop C and less frequently with Tyr254. 
In all these states GABA remains outside the binding pocket identified in the wild-type receptor and does not form any cation-$\pi$ interaction. 
These data suggest that 
Glu204 is crucial to pin the GABA amine group in the most appropriate position to engage
in the cation-$\pi$ interactions typical of Cys-loop pLGICs.

\section*{Conclusion}

We have applied
the enhanced sampling method funnel-metadynamics to investigate the binding mechanism of the neurotransmitter GABA to the wild-type and the most relevant mutant forms of the insect RDL receptor, a prototypical pLGIC. 
This method allows us to overcome the limitations of conventional docking and MD-based techniques by restricting the exploration of the unbound region so to sample a large number of binding and unbinding events within an affordable computational time. 
This leads to a quantitatively well-characterized binding free energy landscape, an accurate estimate of the ligand binding affinity and a detailed description of the atomistic interactions formed by GABA with the RDL receptor in the binding process. 
We disclose the sequence of events that lead GABA to its optimal binding pose in the RDL receptor, starting with the capture from the solvent of its negatively charged carboxylate group by Arg218, followed by the establishment of the crucial interaction with Arg111. Once the carboxylate group is bound to the receptor through Arg111, the positively charged amine group is able to engage with Glu204 by means of a direct or water-mediated hydrogen bond, and with the surrounding aromatic residues through
cation-$\pi$ interactions. Similar interaction patterns are found in other pLGICs.
In our simulations the protein is fully flexible and the waters explicitly represented, allowing us to  investigate the role played
by the protein motion and the solvent during GABA binding.
In particular, we highlight
the functional conformational changes of loop C, which actively participates in the binding process by opening when the ligand approaches the protein
and closing after the binding: this mechanism locks the ligand in the competent binding conformation 
for the activation of the neuroreceptor. 

Simulations on the non-functional Arg111Ala and Glu204Ala mutant forms
demonstrate how
both mutations alter crucial ligand-protein interactions necessary for the binding to the wild-type receptor,
resulting in significant changes in the free energy landscapes and the ligand binding mechanisms.
These changes explain the inactivity of the channel in the mutant forms, in agreement with experimental findings. \cite{Lummis2011,Ashby2012}
Specifically, the mutation Arg111Ala affects the first stages of ligand binding,
preventing GABA from finding a stable anchor point in the complementary subunit and entering the binding pocket with optimal orientation. 
While alternative interactions with residues of the complementary subunit are observed,
they are not sufficient for a stable binding across the two subunits. 
This results in very weak or no binding to the receptor as demonstrated by the free energy landscape and the available experimental data.
In the Glu204Ala mutant form, GABA is able to initially follow the sequence of binding events
found in the wild-type RDL receptor, 
including the interactions with Arg218 and Arg111.
However 
the absence of the negatively charged Glu204, which anchors the GABA amine group in the 
wild-type receptor, prevents 
the completion of the binding process. As a consequence, the neurotransmitter interacts through its carboxylate with Arg111 and the nearby residues of the complementary subunit, while its amine remains unbound without finding suitable interaction partners. 

The possibility of cooperative binding to some or all the five subunit interfaces 
in the RDL receptor ECD
is not accounted for in our work, where the binding of a single ligand to one interface is investigated. In fact the binding of two or three ligands may be necessary to activate 
pLGICs.
\cite{Edelstein1998, Hucho2001}
A cooperative binding mechanism may result into an additive effect for the ligand binding free energy evaluated for each subunit and potential allosteric effects among the subunits that could be worth investigating in the future. 

This study represents the first example where the power of funnel-metadynamics simulations is exploited 
to explore the different binding mechanisms of a ligand towards the wild-type and mutant forms of its molecular target. 
The wealth and accuracy of structural and free energy information provided by our investigation 
are fundamental to complement and interpret the available experimental data.
Similar simulation protocols may be applied to other pLGICs,
whose structures have been recently experimentally resolved, \cite{Hibbs2011, Hassaine2014, Huang2015}
and that shares structural and functional features with the RDL receptor.
The growing availability of structural information from experiments and the improvement and reliability of computational methods like funnel-metadynamics with respect to conventional MD make these studies timely,
since they have the potential to greatly contribute to the detailed understanding of the activation mechanisms of complex and important neuroreceptors.     

\section*{Acknowledgments}
We are grateful for computational support from the UK high performance computing service ARCHER, for which access was obtained via the UKCP consortium and funded by EPSRC grant EP/K013831/1. We also acknowledge the use of the EPSRC UK National Service for Computational Chemistry Software (NSCCS) at Imperial College London (Project ID: CHEM749)
and the Swiss National Supercomputing Center- CSCS (Project ID: s557).
We thank Prof. Sarah Lummis (University of Cambridge) for useful discussions on experimental data. 

\section*{Supporting Information}

A movie showing a favourable path for the binding of GABA to the extracellular domain of the wild-type RDL receptor during the metadynamics simulation. Left: the principal and complementary subunits are shown in gray, loop C in orange and the funnel restraining potential in yellow. GABA and the RDL receptor residues Phe146, Glu204, Phe206, Thr251, Tyr254 in the principal subunit and Tyr109, Arg111, Ser176 and Arg218 in the complementary subunits are explicitly represented.
Right: a zoomed view of the binding process, where GABA and the relevant receptor residues are shown.
This information is available free of charge via the Internet at http://pubs.acs.org/.

\bibliography{abbr,funnel}

\clearpage

\clearpage

\clearpage

\clearpage

\clearpage

\clearpage

\clearpage
\section*{}
\begin{figure*}
\centering
\includegraphics[width=17.8cm]{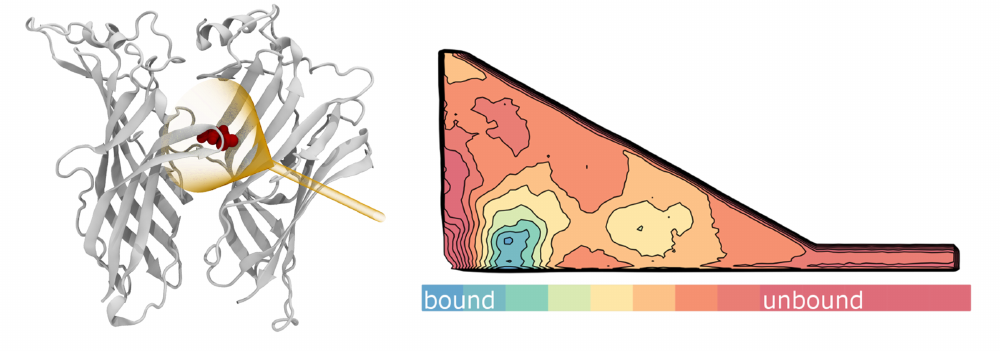}
\caption*{TABLE OF CONTENTS GRAPHIC}
\end{figure*}

\end{document}